
\input harvmac.tex

\input epsf
\ifx\epsfbox\UnDeFiNeD\message{(NO epsf.tex, FIGURES WILL BE
IGNORED)}
\def\figin#1{\vskip2in}
\else\message{(FIGURES WILL BE INCLUDED)}\def\figin#1{#1}\fi
\def\ifig#1#2#3{\xdef#1{fig.~\the\figno}
\goodbreak\topinsert\figin{\centerline{#3}}%
\smallskip\centerline{\vbox{\baselineskip12pt
\advance\hsize by -1truein\noindent{\bf Fig.~\the\figno:} #2}}
\bigskip\endinsert\global\advance\figno by1}

\def \s {\sigma}

\def \ha {\half}
\def \ov {\over}

\def \lr { \lref}

\def\dd {\partial }

\def\l{\lambda}

\def \td {\tilde }
\def \k {\kappa}

\def\n{\noindent}
\gdef \jnl#1, #2, #3, 1#4#5#6{ { #1~}{ #2} (1#4#5#6) #3}

\def\np {  Nucl.  Phys. }
\def \pl { Phys. Lett. }
\def \mpl { Mod. Phys. Lett. }
\def \prl { Phys. Rev. Lett. }
\def \pr  { Phys. Rev. }
\def \cqg { Class. Quant. Grav. }
\def \jmp { J. Math. Phys. }

\lr \bilcal {A. Bilal and C. Callan, \np B394 (1993) 73;
 S. de Alwis, \pl B289 (1992) 278; B300 (1993) 330.}
\lr \cghs {C. Callan, S. Giddings, J. Harvey and A. Strominger,
\pr D45 (1992) R1005. }
\lr \witt {E. Witten, \pr D44 (1991) 314. }
\lr \birdav {N.D. Birrell and P.C.W. Davies, {\it Quantum fields in
curved space} (Cambridge University Press, Cambridge, England, 1982). }
\lr \schutz {B.F. Schutz, `` A first course in general relativity",
(Cambridge University Press, Cambridge, England, 1985).}
\lr \chrisful {S.M. Christensen and S.A. Fulling, \pr D15 (1977) 2088. }
\lr \gidnel {S.B. Giddings and W.M. Nelson, \pr D46 (1992) 2486. }
\lr \hawk {S.W. Hawking, Commun. Math. Phys. 43 (1975) 199. }
\lr \polyakov {A.M. Polyakov, \pl 163B (1981) 207. }
\lr \rustse{H. Verlinde, in
Proceedings of the School in ``String theory and quantum gravity",
Trieste, ICTP (ed. J. Harvey et al) (1991) ;
J.G. Russo and A.A. Tseytlin, \np B382 (1992) 259; A. Strominger,
\pr D46 (1992) 4396.}
\lr \rst {J.G. Russo, L. Susskind and L. Thorlacius,
\pr D46 (1992) 3444; \pr D47 (1993) 533.  }

\lr\horo {G. Horowitz, in Proceedings of the School in ``String theory and
quantum gravity", Trieste, 1992. }

\lr\melv { M.A. Melvin, Phys. Lett. 8 (1964) 65.}

\lr\myers{Y. Kazama, Y. Satoh and A. Tsuchiya,
\pr D51 (1995) 4265;
G. Michaud and  R.C.  Myers, ``Two-dimensional dilaton black holes",
gr-qc/9508063. }
\lr\mignemi{M. Cadoni and S. Mignemi, ``On the conformal equivalence between
2d black holes and Rindler spacetime", gr-qc/9505032.}

\lr\mazzi {D. Mazzitelli and J.G. Russo, \pr D47 (1993) 4490.  }

\lr\harhaw{J.B. Hartle and S.W. Hawking, \pr D28 (1983) 2960.}

\lr\gibb {G.H. Gibbons and K. Maeda, \np B298 (1988) 741.}

\lr\tsey {A.A. Tseytlin, ``Exact solutions of closed string theory",
 hep-th/9505052.}

\lr\das{ S.R. Das and S. Mikherji,  \mpl A9 (1994)  3105 ; T. Chung and
H.~Verlinde,  \np B418 (1994) 305.}

\lr \beken {J.D. Bekenstein, \pr D7 (1973) 2333; D9 (1974) 3292.  }
\lr \hawk {S. Hawking, Commun. Math. Phys. 43 (1975) 199.  }
\lr \thoof { G. 't Hooft,  \np B256 (1985) 727.  }
\lr \hooft { G. 't Hooft, Physica Scripta T36 (1991) 247;
 {\it Dimensional reduction in quantum gravity},
Utrecht preprint THU-93/26, gr-qc/9310006. }
\lr \sussk {L. Susskind,  {\it The world as a hologram},
  preprint SU-ITP-94-33, hep-th/9409089. }
\lr \susugl {L. Susskind and J. Uglum, \pr D50 (1994) 2700.}
\lr \suss {L. Susskind,  {\it Some speculations about entropy in
string theory},    RU-93-44, hep-th/9309145;
J.G. Russo and L. Susskind, \np B 437 (1995) 611; A.~Sen,
{\it Extremal black holes and elementary string states},
  TIFR-TH-95-19, hep-th/9504147; A. Peet, {\it Entropy and supersymmetry of $D$
dimensional extremal electric black holes versus string states},
  PUPT-1548, hep-th/9506200.}
\lr \sredn {M. Srednicki, \prl 71 (1993) 666;
V. Frolov and I. Novikov, \pr D48 (1993) 4545;
D. Kabat and M.J. Strassler, \pl B329 (1994) 46;
C. Callan and F. Wilczeck; \pl B333 (1994) 55.}
\lr \stu {L. Susskind, L. Thorlacius and J. Uglum,
\pr D48 (1993) 3743.  }
\lr \membr {K. Thorne, R. Price and D. MacDonald, {\it Black holes:
the membrane paradigm} (Yale Univ. Press, New Haven, CT, 1986). }
\lr \entropy {E. Keski-Vakkuri and S. Mathur, \pr D50 (1994) 917;
T. Fiola, J. Preskill, A. Strominger
and S. Trivedi, \pr D50 (1994) 3987;  R.C. Myers, \pr D50 (1994) 6412;
J.D. Hayward, DAMTP-R94-61, gr-qc/9412065.
}
\lr \rstf  {J.G. Russo, L. Susskind and L. Thorlacius, \pl B292 (1992) 13.}
\lr \veil  {J.G. Russo, \pr D49 (1994) 5266.}
\lr\wald {R. M. Wald, {\it General Relativity} (University of
Chicago Press, Chicago, 1984).}
 \lr \svv {K. Schoutens, H. Verlinde and E. Verlinde, {\it
Black hole evaporation and quantum gravity},
CERN-TH.7142/94.}
\lr \kant{R. Laflamme and E.P. Shellard, \pr D35 (1987) 2315;
J. Louko, \it{ibid.} 35 (1987) 3760;
J. Louko and T. Vachaspati, \pl B223 (1989) 21;
A. Gangui, F.D. Mazzitelli and M. Castagnino, \pr D43 (1991) 1853.}
\lr\ksacks{R. Kantowski and R. Sacks,\jmp 7 (1967) 443.}
\lr\wein{S. Weinberg, {\it Gravitation and Cosmology},
John Wiley, Inc., New York (1972).}
\lr\cosmol{ M. Hotta, Y. Suzuki, Y. Tamiya and M. Yoshimura,
Progr. Th. Phys. 90 (1993) 689; G. Martin and F.D. Mazzitelli, \pr D50 (1994)
613;
J.S.F. Chan and R.B.~Mann, \cqg 12 (1995) 351;
J. Lidsey, \pr D51 (1995) 6829.}

\baselineskip8pt
\Title{\vbox
{\baselineskip 6pt{\hbox{CERN-TH/95-267}} \hbox{SISSA-ISAS/118/95/EP}
{\hbox{hep-th/9510109}} {\hbox{
   }}} }
{\vbox{\centerline { Soluble models in 2d dilaton gravity}
}}
\vskip -20 true pt
\centerline { A. Fabbri }
\smallskip \bigskip
\centerline {\it SISSA-ISAS and INFN,  }
\smallskip
\centerline {\it Via Beirut 2-4, 34013 Trieste, Italy}
\smallskip
\centerline {\it   fabbri@gandalf.sissa.it}
\bigskip\bigskip\bigskip
\vskip -20 true pt
\centerline  { {J.G. Russo  }}
 \smallskip \bigskip
\centerline{\it  Theory Division, CERN}
\smallskip
\centerline{\it  CH-1211  Geneva 23, Switzerland}
\smallskip
\centerline {\it   jrusso@vxcern.cern.ch}
\bigskip\bigskip\bigskip
\centerline {\bf Abstract}
\bigskip

A one-parameter class of simple  models of two-dimensional dilaton gravity,
which can be exactly solved including back-reaction effects, is
 investigated  at both classical and quantum levels.
This family  contains the RST model as a special case, and it continuously
interpolates between  models having a flat (Rindler)
 geometry and a constant curvature metric with
 a non-trivial dilaton field.
 The processes of formation  of black hole singularities
 from collapsing matter and Hawking evaporation are considered in detail.
  Various physical
aspects of these geometries are discussed, including the
cosmological interpretation.

\medskip
\baselineskip8pt
\noindent

\Date {October 1995}

\noblackbox
\baselineskip 14pt plus 2pt minus 2pt

\vfill\eject
\newsec{Introduction}

  Two-dimensional dilaton gravity models   reproduced the essential features
  of the Hawking model of gravitational collapse, with an exact account
  of back-reaction effects \refs{\cghs,  \rst}.
   An important question that remains is to what extent
  these features are universal or are just   properties peculiar to
    a special model. The CGHS action,
   \eqn\cgh{
   S=  \int d^2 x \sqrt{-g}[ {e^{-2 \phi}(R+
{4 }(\nabla \phi)^2   +4\l^2  )] }\ ,
   }
 is different from the Einstein-Hilbert action restricted to spherically
 symmetric configurations:
$ds^2=g_{ij}(x^i)dx^idx^j+e^{-2\td\phi(x^i) }d\Omega^2 $, $i,j=1,2 $,
  \eqn\einst{
  S_{\rm EH}=  \int d^2 x \sqrt{-g}[ e^{-2 \td \phi}(R+
2(\nabla \td \phi)^2 ) +2  ]  \ ,
  }
 so it is not obvious that the physics of the CGHS model should
be similar to the physics of spherically symmetric Einstein gravity.
  The problem is that the dimensionally reduced Einstein-Hilbert action
  coupled to matter  is not an exactly solvable model.
 It is therefore   important to look for a more general class of
 exactly solvable
 two-dimensional models containing a metric and dilaton field
 in order to have a more universal picture of the dynamics of
 black hole formation and evaporation, at least in the case of spherical
symmetry.
 Several attempts in this direction have been made, either by
 modifying the boundary conditions of ref. \rst , as
 in  ref. \das , or by starting from a different classical action
 (see e.g. ref. \myers ).

 In this paper  we consider a   class of simple solvable models
 whose   classical action is given by
  $S_{cl}=S_0 +S_M$, where
\eqn\uuno{
S_0={1\ov{2\pi}} \int d^2 x \sqrt{-g}\big[ {e^{-{2\ov n}\phi}\big( R+
{4\ov n}(\nabla \phi)^2 \big) +4\l^2 e^{-2\phi}}\big] \ ,
}
and
\eqn\ddue{S_M=-{1\ov{4\pi}}\sum_{i=1}^N \int d^2 x \sqrt{-g}
 (\nabla f_i)^2.
 }
In the case $n=1$ the model will reduce to the  RST model.

The  classical geometries have typically a space-like curvature singularity
with an associated global event horizon, and a   curvature scalar
which goes to zero at spatial infinity. In the frame in which  the
dilaton and metric are   static,
the   generic geometry ($n\neq 1$) does not asymptotically approach   the
Minkowski
geometry, instead it approaches the Rindler metric.
The scale factor goes to zero or to infinity according to whether
 $n>1$ or $n<1$.
 Geometries with non-Minkowskian asymptotic behavior are quite common in
general
theories of 2d dilaton gravity (with a general dilaton potential), and they
also appear in    other contexts,
 such as e.g. ``black strings" in four-dimensional string theory
 (see   \horo \ and refs. therein),
  magnetic flux tubes (e.g. the Melvin vortex in four-dimensional
  Einstein theory \melv ), various (2+1)-dimensional models,
  general gravity   theories with dilaton and Maxwell fields \gibb  ,
etc.\foot{In the simplest
     critical string theory (with zero central-charge deficit) dimensionally
   reduced to two dimensions there are
no asymptotically Minkowskian solutions (the corresponding charged black hole
solutions have a nontrivial asymptotic
  where the scale factor goes to zero).}
  It is therefore of interest to have a simplified context where these
geometries
  can be investigated in detail.



  A basic issue of these types of metrics  is how to define an invariant
   mass in the absence of a preferred asymptotic Minkowski frame.  The standard
   ADM mass is conjugate to the asymptotic Minkowski time.
    For the present models,  the choice of a time scale is somewhat arbitrary
    in that any two time coordinates differing by a multiplicative constant are
     equally valid (there will be, however, a natural  time choice, namely
     the one which, for $n=1$, reduces to the Minkowski time).
    It will be shown here that, once the time coordinate is fixed
 the invariant mass conjugate to this time is conserved in the process
   of black hole formation and evaporation. This quantity constitutes
   a useful parameter  which characterizes the geometry. In particular,
   the zero-curvature ground-state geometry is obtained by setting
   the mass parameter to zero in the general solution.


      A  natural physical application of the models considered here
           is  in the cosmological context (see Section  5). The geometries
corresponding to the cases
      $n>1$, $n<1$, and $n=1$
     are   two-dimensional  analogues of the Robertson-Walker   cosmologies
with parameters
     $k=1$, $k=-1$ and $k=0$, providing   a description
     of expanding or contracting  Universes.


\newsec { Exactly solvable models}

The solvability of the  model of ref. \rst\ is related to the fact
that, after a suitable field redefinition, the action in the conformal gauge
($g_{\pm\pm}=0$, $g_{+-}=-{1\ov 2}e^{2\rho}$) can be written in the
``free field" form \bilcal :
\eqn\free{
S={1\ov{\pi}}\int d^2 x \big[ {1\ov \k }
(-\dd_+\chi \dd_-\chi +\dd_+\Omega\dd_-\Omega) +
\l^2 e^{{2\ov \k }(\chi - \Omega)} + {1\ov 2}\sum_{i=0}^{N}
\dd_+ f_i \dd_- f_i \big] \ ,
}
where
$$
\chi = \k\rho + e^{-2 \phi}-\ha \k\phi  \ ,\ \ \
 \Omega=e^{-2\phi}+ \ha \k \phi\ \ , \ \ \ \  \k={N\ov 12}\ .
 $$

The RST model is not, however, the only dilaton-gravity theory that
can be cast into the   form \free . As we will see below, there are indeed
inequivalent
dilaton gravity models which reduce to the above action upon a field
redefinition.\foot{Field redefinitions involving Weyl scalings do not give
equivalent
theories in dilaton gravity models due to the presence of the   anomaly
term. The matter interacts with the geometry through the conformal anomaly,
which is always constructed in terms of the appropriate physical metric
(for   further discussions on this point see   \rustse ).}

 We would like to find the most general theory whose action can be
 written in the form \free\ and which obeys  the following basic requirements:
\smallskip

\n i) It is reparametrization-invariant. \par \n
ii) It has the correct anomaly term. \par \n
iii) It contains a vacuum solution with $R=0$ as well as asymptotically
 flat solutions.
\par \n
iv) There are no unphysical   fluxes at infinity in the vacuum (in the frame
in which  the metric and the dilaton field are static).
\smallskip
\n Now we will show that the most general transformation that meets the above
requirements is given by
\eqn\Auuu{\chi = \k\rho + e^{-{2\ov n}\phi}+ ({1\ov{2n}}-1)\k\phi }
\eqn\Auuv{\Omega=e^{-{2\ov n}\phi}+ {{\k}\ov{2n}}\phi\ .}
 Condition (i) requires, in particular, that
the cosmological term in eq. \free\  be of the form $\sqrt{-g} f(\phi )=\ha
e^{2\rho} f(\phi)$. The
  most general transformation between
$\chi$, $\Omega$ and $\rho$, $\phi$ satisfying this condition can be
written as
\eqn\Aduea{
\chi = (\k +a)\rho +f_1 (\phi) + g(\rho,\phi) \ ,\ \ \
 \Omega= a  \rho  + g(\rho,\phi)\ .
 }
We can use the freedom to redefine the dilaton field $\phi$
 so as to have
$\chi - \Omega =\k (\rho - \td \phi )$, i.e.
$\k \td \phi =  -f_1(\phi) $ (henceforth $\td \phi=\phi $).
Thus we can write
\eqn\Atrea{\chi =  ( \k+a ) \rho   -\k \phi + g(\rho,\phi) \ ,\ \ \
 \Omega=  a\rho +  g(\rho,\phi) \ . }
Now, in order to obtain the usual anomaly term $-{{\k}\ov{\pi}}
\int d^2x \dd_+\rho \dd_-\rho$,
$g(\rho , \phi )$ must be of the form $g(\rho , \phi)= b\rho+F(\phi)$.
The linear term   $b\rho $ can be reabsorbed into a redefinition
of $a$.
The correct coefficient of the anomaly term is obtained provided
 $(\k+a)^2-a^2 =\k ^2 $\  , i.e.   $a=0 $.
Thus we have $\chi =   \k  \rho   -  \k \phi + F(\phi)  ,\
 \Omega=     F(\phi )$, and we must still   demand
 conditions (iii) and (iv).
 The equations of motion derived from \free\ are
\eqn\qua{
\dd_+\dd_-(\chi - \Omega)=0 \ ,
\ \ \ \ \dd_+\dd_-\chi =-\l^2 e^{{2\ov\k } (\chi - \Omega)}\ .
}
{}From eq. \qua\ one sees that it is always possible to choose a gauge, the
``Kruskal" gauge, where
$\chi=\Omega$. In this gauge
 it is easy to show that the curvature scalar
$R$ is proportional to
\eqn\Acinque{\dd_+\dd_-\rho={1\ov{  F^{'}(\phi)
}}[-\l^2- {F^{''}(\phi)\ov F^{'2}(\phi )} \dd_+\Omega
\dd_-\Omega ]\ .
}
Consider the most general static solutions to   eqs. \qua\
  \rst \ :
\eqn\Asei{
\Omega=\chi=-\l^2 x^+x^- +Q\log (-\l^2 x^+x^-)+
{M\ov{\l  }} \ ,\ \ \ Q,M={\rm const.}
}
Let us first obtain the asymptotic  part of the function $F(\phi )$.
For $(-x^+x^-) \to \infty$, we have (see eq. \Asei )
$\dd_+\Omega \dd_-\Omega \cong -\l^2 \Omega =-\l^2 F(\phi ) $.
 From eq. \Acinque\  we see that there are  zero-curvature
solutions provided
 \eqn\Asette{
 1={{F^{''}}\ov{F^{'2} }}F\ .
 }
 The general solution of eq. \Asette \ is $F(\phi )=ce^{m\phi} $.
The constant $c$ can be removed upon   a proper shift of the dilaton field.
 The presence of the constant $m$
reveals a whole class of new solutions labelled by $n=-{2\ov m}$, with the
vacuum ($R=0$) solution given by
\eqn\Anove{
 e^{2\rho}=e^{2\phi}={1\ov{(-\l^2 x^+x^- )^n}}\ .
}
General configurations approach the vacuum solution in the asymptotic
region.

 Let us note that the condition $R=0$ is   satisfied even if linear terms
 in $\phi$ (which are subleading at infinity and  do not contribute in $F''$)
 are added to $F(\phi )$.    One thus concludes that
 $F(\phi )=e^{m\phi } + B \phi $ is  the most general function $F(\phi)$
 consistent with the existence of zero-curvature solutions.
 In this way  we obtain
\eqn\Aundici{\chi= \k\rho + e^{-{2\ov n}\phi} - (\k -B)\phi \ ,
 \ \ \ \Omega =   e^{-{2\ov n}\phi}+ B \phi
\ . }
The value of $B$ is fixed once   condition (iv) is imposed.
Indeed, consider the constraint equations:
\eqn\Aquattor{
\k t_{\pm}=\k^{-1}(-\dd_\pm \chi\dd_{\pm}\chi
 + \dd_{\pm}\Omega\dd_{\pm}\Omega ) +  \dd_{\pm}^2 \chi +{1\ov 2}
\sum_{i=0}^{N}
\dd_{\pm}f_i \dd_{\pm}f_i \ .}
At infinity we use the $\sigma^{\pm}$ coordinates defined through
$\pm \l x^{\pm} = e^{\pm \l \sigma^{\pm}}$.
In the vacuum   (see \Anove ), $\phi = - {n\ov 2}\l (\sigma^+ -
\sigma^- )$ and $\rho ={{(1-n)}\ov 2} \l (\sigma^+ - \sigma ^- )$ ,
and we get
\eqn\Aquind{
\k t_{\pm}(\sigma^{\pm})=-{{\l^2}\ov 4}[\k  -2n B]\ .
}
In order to have $t_{\pm}(\sigma^{\pm})=0$ in the vacuum,
  $B$
must be equal to ${\k  \ov 2n}$. The most general  model that can be mapped to
the action \free\ obeying conditions (i)-(iv) is thus given by the 1-parameter
class of models
defined by the transformations \Auuu, \Auuv . This leads to the action
\eqn\duno{
S ={1\ov 2\pi} \int d^2 x \sqrt{-g}\bigg[ e^{-{2\ov n}\phi}\big( R+
{4\ov n}(\nabla \phi)^2 \big) +4\l^2 e^{-2\phi}
- \ha \sum_{i=1}^N  (\nabla f_i)^2
 }
$$
+ \k \big( {{1-2n}\ov{2n}}  \phi R
+ {{n-1}\ov{n}}(\nabla \phi)^2- {1\ov 4} R (\nabla^2)^{-1} R \big)\bigg] \ .
$$

\par
\newsec{The classical theory}
 Let us first consider  the classical theory $\hbar  \to 0 $.
Once $\hbar  $ is restored in the formulas,   the last
 three terms in eq. \duno\ go away in this limit, and we are left with action
\uuno .
The equations of motion derived from this action are
\eqn\dtre{
g_{\mu\nu}\big[ {4\ov n}\big( -{1\ov 2}+{1\ov n}\big) (\nabla \phi)^2 -{2\ov n}
\nabla^2 \phi - 2\lambda^2 e^{{{2-2n}\ov n}\phi}\big] +
{4\ov n}\big( 1-{1\ov n}\big) \dd_{\mu}\phi \dd_\nu \phi
}
$$
+{2\ov n}
\nabla_{\mu}\dd_{\nu}\phi + e^{{2\ov n}\phi}T_{\mu\nu}^M =0\ ,
$$
\eqn\dquattro{
{R\ov n} - {4\ov {n^2}}(\nabla \phi)^2
+{4\ov n}\nabla^2\phi +4
\lambda^2 e^{{{2-2n}\ov {n}}\phi}=0\ ,
}
\eqn\dcinque{
\nabla^2 f_i =0\ .
}
Equation \dtre\ results from the variation of the metric and \dquattro\ is the
dilaton equation of motion.
 In the conformal gauge  $g_{\pm\pm}=0$, $g_{+-}=-{1\ov 2}
e^{2\rho}$   the equations of motion  become
\eqn\dotto{
-{4\ov {n^2}}\dd_+\phi\dd_-\phi+ {2\ov n}\dd_+\dd_-\phi
- \lambda^2 e^{{{2-2n}\ov n}\phi + 2\rho} =0\ ,
}
\eqn\dnove{
{2\ov n} \dd_+\dd_-\rho +{4\ov{n^2}}\dd_+\phi\dd_-\phi
-{4\ov n}\dd_+\dd_-\phi +\lambda^2 e^{{{2-2n}\ov n}\phi+
2\rho} =0\ ,
}
\eqn\ddieci{\dd_+\dd_- f_i =0\ ,
}
and the constraints
\eqn\dundici{
e^{-{2\ov n}\phi}
\big[ {4\ov n}\big( 1-{1\ov n}\big) \dd_{\pm}\phi\dd_{\pm}\phi +
{2\ov n}\dd_{\pm}^2\phi -{4\ov n}\dd_{\pm}\rho \dd_{\pm}\phi \big]
- {1\ov 2} \sum_{i=0}^{N}\dd_{\pm}f_i \dd_{\pm}f_i =0\ .
}
{}From eqs. \dotto\ and \dnove\ it follows that
\eqn\ddodici{
{2\ov n}\dd_+\dd_- (\rho - \phi)=0\ ,
}
 i.e. $\rho=\phi+f_+(x^+)+f_-(x_-)$. It is always possible to perform
a coordinate transformation $x^{\pm} \to x^{\pm '}=f(x^{\pm})$ ,
which preserves the conformal gauge and for which $\rho=\phi$.
 In this (Kruskal-type) gauge the remaining equations take the form
 \eqn\dtredici{
 \dd_+\dd_-(e^{-{2\ov n}\phi})=-\lambda^2 \ ,\ \ \
\dd_{\pm}^2(e^{-{2\ov n}\phi})=
-{1\ov 2}\sum_{i=0}^{N}\dd_\pm f_i\dd_\pm f_i \ ,
}
so that the general solution is given by
\eqn\dqui{
e^{-{2\ov n}\phi}=e^{-{2\ov n}\rho}=-\lambda^2 x^+x^- +h_+(x^+) +
h_-(x_-) \ ,
}
where $h_\pm (x^\pm )$ are arbitrary functions of $x^\pm $
subject to the constraints \dundici .

\subsec{Static solutions}

In the Kruskal gauge the general static solution is given by
 (see eq. \dqui )
\eqn\dsedici{
e^{-{2\ov n}\phi}=-\lambda^2 x^+ x^- +Q\log (-\lambda^2 x^+x^-) +
{M\ov{\l}} \ ,
}
 i.e. for these solutions there exists a time-like
Killing vector at infinity representing time translation invariance
with respect to the time coordinate $t$, where $t=\ha \log {x^+\ov x^-} $
(see also below).  In   subsection 3.2 it will be shown that $M$ can be
interpreted as
the mass of the black hole.
 The parameter $Q$ represents a uniform   (incoming and
 outgoing) energy density flux. Indeed,
    the constraint equations \dundici\  applied to the solution
\dsedici\ give $T_{\pm\pm}={Q\ov{x^{\pm 2}}} $ or,
 introducing
$(\sigma^{\pm})$ defined by
$\lambda x^{\pm}=\pm e^{\pm \lambda \sigma^{\pm}}$,
  $T_{\pm\pm}= \l^2 Q$.

Let us consider the static solution with $Q=0$ :
\eqn\tuno{
ds^2=-{1\ov{({M\ov{\l}}-\lambda^2 x^+x^-)^n}}dx^+dx^- \ ,\ \ \
 e^{-{2\ov n}\phi}={M\ov{\l}}-\lambda^2 x^+x^-\ .
}
The corresponding curvature scalar $R$ is given by
\eqn\tdue{
R=8e^{-2\rho}\dd_+\dd_-\rho =4M\lambda n
\bigg[{M\ov{\l}}-\lambda^2 x^+x^-\bigg]
^{n-2}.
}
Consider   the range $0<n<2$ .\foot{When $n>2$ the geometry is very different;
  for simplicity  here this case will be excluded from the discussion.}
In this case we get the standard picture of the $n=1$ solutions,
i.e. a space-like singularity located at $x^+x^-={M\ov{\lambda^3}}$
and an asymptotically flat region   for  $-x^+x^- \to \infty$
($x^+\to \infty$ defines the future null infinity $I_R^+$ and
$x^-\to -\infty$ stands for the past null infinity $I_R^-$).
The event horizon is at $x^-=0$. The Penrose diagram is
identical to the standard $n=1$ case (see e.g. \cghs   ).

{}From eq. \tdue\ we see that for $n=0$ the two-dimensional spacetime is flat.
This is not, however, a trivial solution, since the coupling constant
$e^{2\phi }$ is non-trivial and it becomes singular on a space-like line.
To take the limit $n\to 0$  we must first
rescale the dilaton field $\phi\to \td\phi= n\phi $. The
classical action \uuno\ takes the simple form
$$
 S_0={1\ov 2\pi }\int d^2x \sqrt{-g} \big( e^{-2\td \phi } R + 4\l^2 \big)\ .
$$
This is precisely what one gets from the CGHS action \cgh\ if the
metric is redefined by $g_{\mu\nu }\to e^{2\phi}g_{\mu\nu }$.
The case $n=0$ represents an unconventional black hole in the
sense that there is a space-like singularity in the coupling (and hence a
horizon),
 but the two-dimensional curvature vanishes (for a recent discussion on
this model, see ref. \mignemi ).\foot{
 In the dimensional reduction interpretation, the singularity in
$e^{2\td\phi }$ translates into a curvature singularity of
the four-dimensional metric
$ds^2=g_{ij}dx^idx^j+e^{-2 \td \phi}d\Omega^2$.}

For $n=2$  the two-dimensional curvature is constant. However, the same
considerations
as for the case $n=0$ apply:  the dilaton field is singular on a
space-like line and the full geometry still has a black-hole interpretation,
with an event horizon at $x^-=0$. In Section  4 we will see that at the quantum
level the curvature of the $n=2$ model is no longer constant, and it becomes
singular on a curve where the coupling reaches some  finite critical value.
\par

  Let us now perform the coordinate transformation
($x^+$,$x^-$) $\to$ ($\sigma$,$t$) by means of the relation
$\pm\l x^{\pm}=f(\l\sigma)
e^{\pm  \lambda t }$, where
$f$ is a generic function of $\l\sigma$.
In this new coordinate system the line element and dilaton field take the form
\eqn\ttre{
ds^2 = {1\ov{({M\ov{\l}}+f^2(\l\sigma))^n}}
(-f^2(\l\sigma)dt^2 + f^{'2}(\l\sigma)d\sigma^2)\ ,\ \ \
\phi = -{n\ov 2} \log \big[ {M\ov {\l}}+ f^2 (\l\sigma )\big] \ .
}
 A convenient coordinate system that will be used here
 is $f(\l\sigma)=e^{\l\sigma}$,
 \eqn\tquat{
 ds^2= {{e^{2(1-n)\l \sigma}}\ov{(1+{M\ov{\l}}e^{-2\l \sigma})^n}}
(-dt^2 + d\sigma^2) \ ,\ \ \
 \phi =-{n\ov 2}\log \big[{M\ov {\l}}+e^{2\l\sigma} \big] \ .
 }
This coordinate system is suitable to calculate the mass of the black hole
by means of the ADM procedure (see Section 3.2).
{}From eq. \tquat\ we see that  the metric does not asymptotically
approach the Minkowski metric unless $n=1$.  Instead we  observe the remarkable
fact that for any $n\neq 1 $ the geometry approaches
 the Rindler metric. Indeed, consider first
 the vacuum solutions (i.e. with $M=0$)
in terms of the spatial coordinate $x$ defined by ${{df}\ov{f^n}}=
\l dx$, that is
\eqn\tcinque{f^{1-n}= \l (1-n)x\ , \   n< 1\ ,\ \ \ \ \ \ \
f^{1-n}= \l (n-1)(x_1-x) \ , \  n> 1\ ,}
where $x_1$ corresponds to the point $f=\infty $. In this frame we get,
e.g. for $n<1$,
\eqn\tsei{
ds^2=dx^2 - [\l(1-n)x]^2 dt^2\  ,\ \ \
\phi =-{n\ov{(1-n)}}\log [ \lambda (1-n)x ] \ ,\ \
}
 that is, the Rindler metric. In the special case $n=1$  one obtains $f=e^{\l
x}$ and the geometry is
the familiar linear dilaton vacuum, i.e. the Minkowski metric
$ds^2= - dt^2+dx^2$ and $\phi=-\l x$.
\par

For $M\neq 0$ we have
\eqn\tnove{
ds^2 = dx^2 - F(\l x) dt^2 \ ,\ \ \ \
 F(\l x)={{f^2}\ov{({M\ov{\l}}+f^2)^n}}\ ,
}
\eqn\totto{
{{df}\ov{({M\ov{\l}}+f^2)^{n/2}}}=\l dx \ .
}
 Although it is not possible to integrate \totto\ in a closed form for generic
 $n$ (in the case $n=1$  eq. \totto\  gives $F(\l x)=\tanh^2 (\l x)$\ ),
   the geometry can   be visualized by examining
the form of $F(f^2 )$.
Near the horizon, $f\cong 0$ and
$F(f^2)\cong \big( {\l\ov M}\big)^n f^2 \cong 0$.
In the asymptotic region, $f\to \infty $ and $F(f^2)=f^{2-2n}$.
For $n<1$ the ``cigar"     expands, $F\to \infty $, and for $n>1$ it shrinks
 (see also Section 5 and figs. 1, 2, 3 therein).
In going to the $x$ coordinates, when $n>1$ the point $f=\infty $ is mapped
into
a finite point $x_1$, since
  $f^{1-n }\sim \l (n-1) (x_1-x) $ and
  $F(\l x)\sim [\l (1-n) (x_1-x)]^2$.

The fact that on the horizon $F(\l x)\sim (\l x)^2$ for all $n$ shows
 that the Hawking temperature
  will be given by $\l/2\pi $ , irrespective of the value of $n$. This result
is unambiguous once the time
 scale is fixed, and it will be confirmed below by means of
 two alternative derivations.

\subsec {ADM mass}
In this paragraph we perform the calculation of the ADM mass for these
generalized black-hole configurations. We stress once again that
in the absence of a (preferred) asymptotic Minkowski time, there is no
unique possible definition
of `mass'. The   calculation that follows corresponds to the
mass   conjugate to the time $t$ introduced before;  this is a natural time
choice in that   it reduces to the Minkowski time for the $n=1$ model.
The  introduction of this mass parameter   is useful since
it is   a conserved quantity  in the process of evaporation characterizing the
geometry (see below).

If we denote by $A_{\mu\nu}$ the
gravitational field equations and by $\xi^{\mu}$ a Killing vector
field, then $j_{\mu}=A_{\mu\nu}\xi^{\nu}$ should be a conserved current
and the  corresponding conserved charge density    a total
divergence.
The corresponding charge is determined as a surface term
at infinity.
 In the case $\xi^{\mu}=(1,0)$, representing time translation invariance,
the only conserved quantity is the total energy or mass.

We work in the ($\sigma$,$t$) coordinate system introduced  before.
In this frame the metric \tquat , which for the moment we write generically
as $ds^2=-e^{2\rho}(dt^2 - d\sigma ^2)$, and the dilaton depends only on
$\sigma$.
The ${00}$ component of eq. \dtre\ now reads
\eqn\Buno {
A_{00}=e^{-{2\ov n}\phi}g_{00}g^{11}[{4\ov n}(-{1\ov 2}+{1\ov n})
(\dd_1\phi )^2-{2\ov n}\dd_1^2\phi + {2\ov n}\dd_1\rho\dd_1\phi]
-2 g_{00}\l^2 e^{-2\phi} \ .
}
In the linear  approximation $A_{00}$ is good enough to prove the
conservation of the charge.
Let us expand $\rho$ and $\phi$ around their vacuum values, i.e. $\phi=
-n\l\sigma +\delta\phi$ and $\rho=(1-n)\l\sigma +\delta \rho$.
Note that $\delta \phi=\delta\rho$ (see \tquat ), so that the last term in
\Buno\ gives no first-order contributions. Using also  $g_{00}g^{11}=-1$,
 we find
\eqn\Bdue{
j_0=e^{2\l\sigma}[{2\ov n}\dd_1^2\delta\phi +{6\ov n}\l\dd_1\delta\phi
+{4\ov n}\l^2 \delta\phi ]=\dd_{\sigma}[e^{2\l\sigma}({2\ov n}
\dd_1\delta\phi +{2\ov n}\l\delta\phi )] \ .
}
This means that
\eqn\Btre{
\int d\s j_0=[e^{2\l\sigma}({2\ov n}\dd_1\delta\phi +{2\ov n}\l\delta\phi )]
\bigg|_{\sigma=\infty}\ .
}
Now let us explicitly determine  $\delta\phi$. From
$$
e^{-2\phi}=[{M\ov{\l}}+e^{2\l\sigma}]^n=e^{2\l n\sigma}
[1+{M\ov{\l}}e^{-2\l\sigma}]^n\sim e^{2\l n \sigma}[1+n{M\ov{\l}}
e^{-2\l\sigma}]\ ,
$$
we get $\delta\phi=-{{nM}\ov{2\l}}e^{-2\l\sigma}$.
Substituting in eq. \Btre , we finally obtain
\eqn\Bquattro{\int d\s j_0=M \ .}

\subsec{ Dynamical formation of black holes}
Let us now return to the general solution, eq. \dqui  ,
  and consider the problem of dynamical
black-hole formation starting from the vacuum.
The discussion is a straightforward generalization of the $n=1$ model
of ref. \rst , but we briefly outline it in order to fix the notation.

Using the constraints (see eq.  \dtredici )  we can express the
general solution in terms of physical quantities, such as
the Kruskal momentum and energy \rst
$$
P_+(x^+)=\int_0^{x^+}dx^+T_{++}^M(x^+) \ ,\ \ \
M(x^+)=\lambda \int_0^{x^+} dx^+ x^+ T_{++}^M(x^+) \ .
$$
We get
\eqn\quno{
e^{-{2\ov n}\rho}=e^{-{2\ov n}\phi}= -\lambda^2 x^+(x^- +
 \l^{-2} P_+(x^+) ) +\l^{-1} M(x^+)\ .
}
The  corresponding    curvature scalar $R$  is given by
\eqn\qdue{
R=4\lambda n M(x^+)\bigg[ {{M(x^+)}\ov{\l}}
-\lambda^2 x^+(x^- + \l^{-2} P_+(x^+) )\bigg] ^{n-2}\ .
}
It exhibits a singularity located at
\eqn\qtre{
M(x^+)-\lambda^3 x^+\big( x^- + \l^{-2} P_+(x^+) \big) =0.
}
There is an apparent horizon at $x^- +\l^{-2}P_+(x^+)=0$ and
an event horizon at $x^- +\l^{-2}P_+(\infty )=0$.

Consider the case of an incoming shock-wave at $x^+=x_0^+$ represented by
the stress tensor $T_{++}^M={1\ov 2}
\sum_{i=0}^{N}\dd_+ f_i \dd_+ f_i=a\delta (x^+ - x_0^+)$.
The constraint equation is then easily satisfied by
\eqn\qquattro{e^{-{2\ov n}\phi}
=e^{-{2\ov n}\rho}=-a(x^+ - x_0^+)\theta (x^+ - x_0^+)
-\lambda^2 x^+x^- .}
In the region $x^+<x_0^+$ the geometry is that of the vacuum,
whereas in the region $x^+>x_0^+$ the geometry is that of the static
black-hole configuration discussed previously (with   mass
parameter $M=ax_0^+ \l$).

\newsec{Quantum theory}
 \subsec{Hawking radiation}
 Let us first discuss the
 Hawking radiation ignoring back-reaction effects due to the
 evaporation.
We consider the   quantization of  the $N$ massless scalar
fields in the fixed background of a black hole formed by the collapse
of an incoming shock-wave.
Since the $f_i$'s are free fields, they admit the decomposition
$f_i=f_{iL}(x^+)+
f_{iR}(x^-)$, where $f_{iL}$ represents the incoming wave and
$f_{iR}$ the outgoing one.

In the region $x^+ >x_0^+$ the metric is given by
\eqn\cdiciotto{
e^{-{2\ov n}\rho}=-\l^2 x^+ (x^- + {a\ov{\l^2}})
+ax_0^+  \  .
}
Consider the frame ($\sigma^+_{\rm out}$, $\sigma^-_{\rm out}$), appropriate
 for an out observer, defined by
\eqn\cdiciannove{\l x^+ =e^{\l\sigma^+_{\rm out}}\ , \ \ \
 -\l (x^- + {a\ov{\l^2}})=e^{-\l\sigma^-_{\rm out}}\ .
 }
 The  ``in" vacuum   $|0\rangle _{\rm in}$ is defined as being annihilated by
 the negative frequency modes  with respect to the ``in" time
$(\sigma^{+} _{\rm in},\sigma^{-} _{\rm in})$, $\pm\l x^{\pm}=
e^{\pm \l \sigma^{\pm}_{\rm in}}$.
The Hawking radiation will be determined as usual in terms of the
Bogolubov transformation between the ``in" and ``out" coordinate systems.
Since this
is independent of $n$ the calculation is formally identical to the case $n=1$
so it will not be reproduced here (see  e.g. ref. \gidnel).
One obtains
 \eqn\cquar{
{}_{\rm in}\langle T_{--} \rangle_{\rm in} =
{{N\lambda^2}\ov{48}}\bigg[ 1- {1\ov{(1+{a\ov{\lambda}}e^{\lambda
\sigma^-_{\rm out}})^2}}\bigg] \ .
}
Near the horizon, $\s^-_{\rm out}\to \infty $, and ${}_{\rm in}\langle
T_{--}\rangle _{\rm in}$ approaches the constant
value ${{N\l^2}\ov{48}}$. In this region it can be shown that
 ${}_{\rm in}\langle N^{\rm out}_w \rangle _{\rm in} \sim {{e^{-{{2\pi w }
\ov{ \l}}}}\ov{1- e^{-{{2\pi w}\ov{\l}}}}}$ (where $N_w^{\rm out}$ is the
number operator of the out modes of frequency $w$), that is,
     the outgoing flux of radiation is thermal
at the Hawking temperature $T_H ={{\l}\ov{2\pi}}$.

\subsec{Back-reaction}

 The inclusion of exact back-reaction effects can be done as in the $n=1$
 case \rst , by solving the semiclassical equations of motion corresponding
 to the effective action including one-loop effects, eq. \duno .
In terms of $\chi ,\Omega $ the   mathematics is identical
to the $n=1$ case. However, some physical quantities, such as e.g. the
curvature scalar and the dilaton, have an $n$-dependent time evolution.
 Here we will just point out  the general features and the main
differences with respect to the standard $n=1$ case.

In terms of $\chi, \Omega $ the vacuum solution is
\eqn\scinque{
\Omega=\chi=-\l^2 x^+x^- -
{{\k}\ov 4}  \ln{(-\l^2 x^+x^-)} \ .
}
 The general time-dependent solution that describes
 the collapse of general incoming massless matter and subsequent evaporation
is given by :
\eqn\ssette{\Omega=\chi=-\l^2 x^+
\big( x^- + \l^{-2} P_+(x^+) \big) -{{\k}\ov 4}\log (-\l^2 x^+x^- )
+\l^{-1} M(x^+)   \ .
}
The curvature scalar of the corresponding geometry is
\eqn\seuno{
R = 4ne^{-2\rho} {1\ov [e^{-{2\ov n}\phi}   - {\k\ov 4} ] }
(\l^2 + {4\ov {n^2}}\dd_+\phi\dd_-\phi e^{-{2\ov n}\phi}) \ .
}
We notice that there is a singularity
along the line $\phi=\phi_{\rm cr}=-{n\ov 2}\log {{\k}\ov 4}$.
This line turns out to be time-like
 if $T_{++}< {{\k}\ov{4x^{+2}}}$, and it becomes space-like
  as soon as $T_{++}> {{\k}\ov{4x^{+2}}}$ .

Note that in the $n=2$ case, which classically corresponded to a constant
curvature,  the geometry has undergone an important change: once the one-loop
effect
has been incorporated, not only is the curvature   not  constant, but it blows
up at $\phi=\phi_{\rm cr}$. This is not a surprise; while at the classical
level the curvature was constant, the coupling became strong in a certain
region. The quantum-corrected metric   approaches   the classical (constant
curvature)
metric asymptotically, but it departs from it in the strong coupling region.

As  in the $n=1$ case
it is always possible to impose boundary conditions on the
time-like singularity such that the curvature remains finite
 (in this picture the critical line can be viewed   as a boundary
of the space-time, just as   the line $r=0$ in the spherically symmetric
reduction of 4d Minkowski space).
 Since the denominator vanishes on the singularity,
  the curvature will remain finite   only if
\eqn\sedue{
\l^2 =-{4\ov{n^2}}\dd_+\phi\dd_-\phi e^{-{2\ov n}\phi}
\mid_{\phi=\phi_{cr}}\ ,
}
or $(\nabla\phi )^2=n^2({\k\ov 4})^{n-1} \l^2$.
This can be accomplished
by demanding $\dd_+\Omega \mid_{\phi=\phi_{\rm cr}}=
\dd_-\Omega \mid _{\phi=\phi_{\rm cr}}=0 .$
 It should be remembered
that in order to take the limit $n\to 0$, one must first rescale
$\phi\to n\phi $.

Energy conservation can be checked just as in the case $n=1$ \rst .
  We must compute the quantity
\eqn\seundici{
E_{\rm out}=-\ha \l \int_{-\infty}^{x_1^-} dx^-
\big[ x^- + \l^{-2} P(x_1^+) \big] \sum_{i=0}^{N} \dd_- f_i\dd_-f_i  \ ,
}
where $x^+_1$ represents the advanced time at which the incoming energy flux
stops, and $x^-_1=-\l^{-2}  P(x_1^+)$.
The result of the integration exactly reproduces   the total incoming energy.

\newsec{ Cosmological models}

The theory \uuno\ is similar to the dimensional reduction of four-dimensional
Einstein gravity \einst\ with
 \eqn\uppo{
ds^2=g_{ij}(x^i) dx^i dx^j + e^{-{2 \td \phi (x^i)}}d\Omega^2\ , \ \ \
\td\phi=\phi/n \ .
}
 For $n<2$
the general   model \uuno\  with the change $\l^2\to -\l^2 $
exhibits   interesting cosmological solutions, which may be regarded as  toy
Kantowski-Sacks models \ksacks, describing `spatially
homogeneous' spacetimes with general line element
\eqn\uppa{
ds^2 = A (t) (-dt^2 +d\sigma^2) +B (t) d\Omega^2 \ .
}
$A(t)$ and $B(t)$ are generic functions of $t$
 (a discussion in the
case of the $n=1$ model can be found in ref. \mazzi ;   other generalized
dilaton gravity models are discussed in  ref. \cosmol ).

Consider a homogeneous distribution of conformal matter,
with
$$
T_{++}^{\rm matter}=T_{--}^{\rm matter}={1\ov {4}}\sum_{i=0}^{N} (\dd_t f_i )^2
=c \ .
$$
In the conformal gauge,
the scale factor and dilaton of the  homogeneous solution
$ds^2=e^{2\rho (t)} (-dt^2+d\s^2)$ have the following general  form
\eqn\ppqq{
e^{2\rho}= (e^{2\l t }+{c\ov\l}t+m )^{-n} e^{2\l t}\ ,
\ \ \ \ e^{ {2\ov n}\phi }=\big( e^{2\l t} + {c\ov\l} t + m \big)^{-1}\ .
}
When either $c\neq 0$ or $m<0$  the evolution starts at some finite
$t=t_0$ where (for $n>0$) $e^{2\phi}$, $e^{2\rho}$
and $R$ are all  singular. When
$m>0$ and $c=0$ the evolution starts at $t=-\infty $, where
the solution is regular.
The behavior at $t\to\infty $
does not depend on the values of $m$ and $c$,
as    is clear from eq. \ppqq .

Let us discuss in detail  the simplest case $m>0$ and $c=0$.
 The scale factor is
\eqn\ppp{
e^{2\rho}= (e^{2\l t }+m )^{-n} e^{2\l t}.
}
The curvature and the dilaton field are given by
\eqn\rccc{
 R= -4n \l^2 m (e^{2\l t}+m)^{n-2}\ ,\ \ \
 e^{2 \phi}=(e^{2\l t}+m)^{-n} \ .
 }
{}From \ppp\ we note that the   coupling $e^{{2\ov n}\phi}$   always stays
finite.
For $t\to -\infty$,  $e^{2\rho (t)}\to 0$,
$R\to 4n \l^2  m^{n-1}$ and $e^{{2\phi\ov n}}\to m^{-1}$.
Then the  Universe begins to expand and the subsequent evolution will be
dictated
by
the value of $n$.
As   $t\to \infty$ we have
\eqn\cptt{
e^{2\rho}\to   e^{2\l (1-n) t}\ ,
}
i.e. the scale will increase for $n<1 $ and decrease for $n>1$,  whereas
$e^{{2\ov n}\phi}\to 0$ and
 $R\to 0$ irrespective of the value of $n$.
 Thus for $n<1$ the  Universe is open and expands forever (Fig.  1),
 in the case $n=1$ the expansion
slows down to zero asymptotically (Fig.  2),   and   for $n>1$ the  Universe is
closed:
 at a certain time the expansion
stops  and the  Universe begins to contract (see Fig.  3).
It is interesting to note that in this last case the collapse takes place
in   a finite proper time, and in
the weak coupling region where $e^{{2\ov n}\phi}=0$.
\par

\ifig\fone{Euclidean embedding of the metric for $0<n<1$. In the case
$n=0$ the metric is that of the plane. For $n<0$ the Euclidean embedding does
not exist: the geometry
describes a hyperbolic  Universe that cannot be represented as a
two-dimensional surface in  three -or higher- dimensional Euclidean space.}
 {\epsfxsize=7.5cm \epsfysize=4.5cm \epsfbox{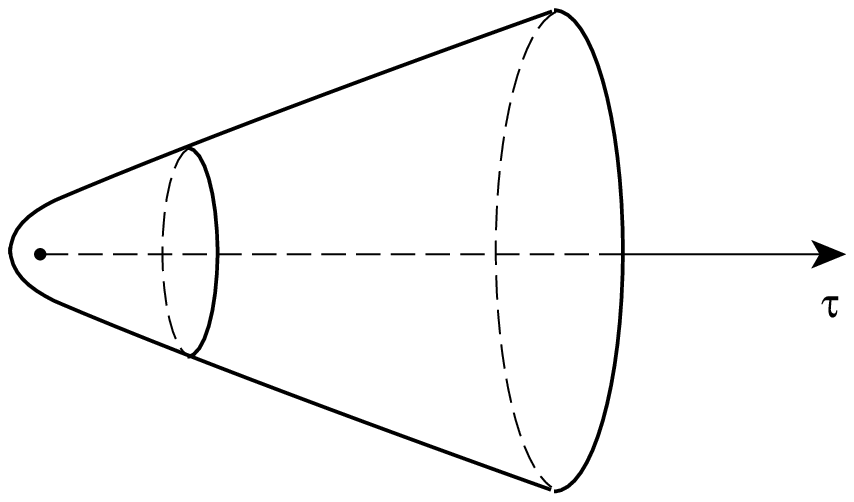}}

\ifig\ftwo{Standard ``cigar" geometry for $n=1$.}
{\epsfxsize=7.5cm \epsfysize=2.4cm \epsfbox{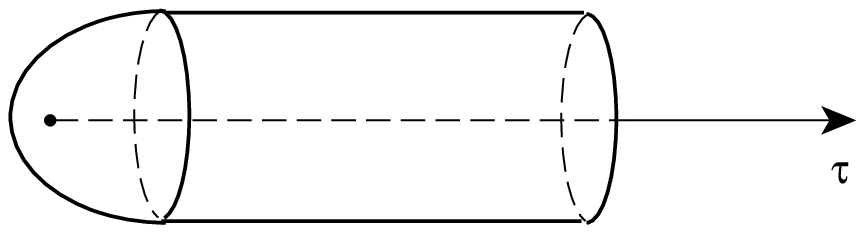}}

\ifig\fthree{ Euclidean embedding of the metric for $1<n<2$.
For $n=2$ the euclidean metric reduces to the metric of the sphere.}
{\epsfxsize=7.5cm \epsfysize=3.1cm \epsfbox{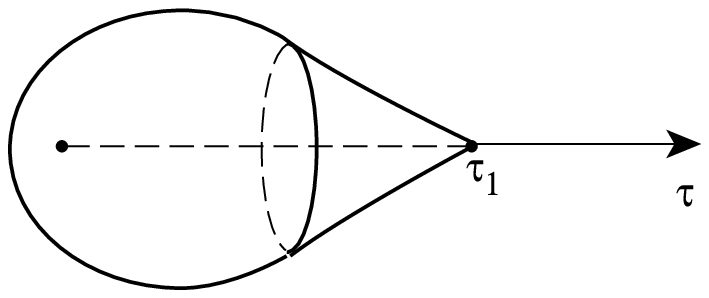}}

Let us introduce the cosmological time $\tau $ and
consider the Euclidean  metric
\eqn\eucli{
ds^2=d\tau ^2 +F(\l\tau )dx^2\ ,\ \ \
F(\l\tau )={e^{2\l t} \ov (m+e^{2\l t})^n }\ ,
\ \ {e^{\l t} dt\ov (m +e^{2\l t})^{n/2} } = d\tau \ .
}
The compact space coordinate $x$  must have period
$2\pi/\l $ in order for the metric to be free from   conical singularities at
$\tau=0 $.
This is clear from the fact that in the region $\tau\cong 0$ ($t\to -\infty $)
 one has $F(\l\tau )\cong \l^2\tau^2 $.
The metrics with $1<n<2$ will then have a conical singularity at
$t=\infty $ (Fig.  3). Indeed, at $t\to \infty $,  one finds
$F(\l\tau )=\l^2(n-1)^2(\tau _1- \tau)^2 $, which   implies a conical
singularity at $\tau=\tau_1$ \
($t=\infty $) with deficit angle
equal to $2\pi (2-n)$.
For $n=2$  there is no conical singularity  and the metric is that of the
sphere, $F(\l \tau )= \sin^2 (\l \tau )$.

Thus the simplest cosmologies with $c=0$ and $m>0$ contain
expanding and contracting (non-``isotropic")  Universes with no initial
singularities.
 In the case $n>1$ the  Universe recollapses in spite of the absence of
 matter energy density ($c=0$). This is due to the fact that  for $n>1$,
as the weak limit $e^{{2\ov n}\phi }\to 0 $ is approached ($t\to \infty $),
 the scale factor $e^{2\rho }$ must go to zero in order to compensate the
increase of the  cosmological term  in the action
(see eq.   \uuno ).


A final remark concerns the   case   when $n=-1$. For this model,
 in the gauge $\rho=\phi $, the functions (see eq. \uppa )
$A=e^{2\rho }$ and $B=e^{-{2\ov n}\phi}$ become
 the same. An homogeneous solution in this gauge is
$ds^2=(m+2 \l^2 x_0^2)(-dx_0^2+dx_1^2) ,\  e^{2\phi}= m + 2  \l^2 x_0^2$.
The four-dimensional metric \uppo\ can thus be written as
 \eqn\pupp{
ds^2=-d\tau^2 + R^2 (\tau) (d\sigma^2  + d\Omega^2 )\ ,\
}
which means (in this four-dimensional interpretation)
that the spatial section of the metric remains constant throughout the
evolution.
 For large $\tau $, the radius of the  Universe increases as
$R(\tau )\sim \sqrt{\tau} $ so that the
Hubble constant  $H\equiv {1\ov R} {dR\ov d\tau }$ goes to zero as
${1\ov \tau }$.
 This is quite satisfactory, since the behavior $H\sim {1\ov R^2 }$ is
characteristic of standard   radiation-dominated
Friedmann-Robertson-Walker ($k=0$) cosmologies.


\bigskip\bigskip

\noindent $\underline{\rm Acknowledgements}$:
 We would like to thank D. Amati for useful discussions. J.R. thanks SISSA,
Trieste, for hospitality during the course of this work.

  \listrefs
\vfill\eject
 \end